\documentclass{pasa}%
\pdfoutput=1

\title[TGSS-RSADR1]{A Rescaled Subset of the Alternative Data Release~1 of the TIFR GMRT Sky Survey}
\author[Hurley-Walker, N.]{Hurley-Walker, N.$^1$\thanks{nhw@icrar.org}\\
\affil{$^1$International Centre for Radio Astronomy Research, Curtin University, Bentley, WA 6102, Australia}}%
\jid{PASA}
\doi{10.1017/pas.\the\year.xxx}
\jyear{\the\year}

\usepackage[authoryear]{natbib}
\bibpunct{(}{)}{;}{a}{}{,}
\usepackage{amssymb} 
\usepackage{aas_macros}
\usepackage{hyperref}

\hypersetup{colorlinks,citecolor=blue,linkcolor=blue,urlcolor=blue}
\usepackage{rotating}
\def\Fig{Figure}

\def\Sect{Section}

\def\Tab{Table}

\newcommand{\farcm}{\mbox{\ensuremath{.\mkern-4mu^\prime}}}
\newcommand{\fdg}{\mbox{\ensuremath{.\!\!^\circ}}}
\newcommand{\arcdeg}{\ensuremath{^{\circ}}}

\begin{document}%
\begin{abstract}
This Rescaled Subset of the Alternative Data Release 1 to the Tata Institute of Fundamental Physics Giant Metrewave Radio Telescope Sky Survey (TGSS-RSADR1) modifies the initial data release of TGSS-ADR1 \citep{2017A+A...598A..78I} to
bring that catalogue to the same flux density scale as the extragalactic catalogue from the GaLactic and Extragalactic All-sky Murchison Widefield Array survey \citep[GLEAM;][]{2015PASA...32...25W,2017MNRAS.464.1146H}. In this paper we motivate the derivation of correct and complementary flux density scales,
introduce a methodology for correction based on radial basis functions, apply it to TGSS-ADR1, and create a modified catalogue, TGSS-RSADR1. This catalogue comprises 383,589~TGSS-ADR1 sources with updated flux density and flux density uncertainty values, and covers $\mathrm{Declination}\leq+30^\circ$, $|b|\geq10^\circ$, a sky area of 18,800\,deg$^2$.
\end{abstract}
\maketitle%
\section{INTRODUCTION }
\label{sec:intro}

Wide-area surveys are extremely useful for large-scale studies of the properties of astrophysical objects, with a host of new surveys being conducted by Square Kilometer Array (SKA) pathfinders (see \cite{2013PASA...30...20N} for a review). Radio surveys probe the populations of active galactic nuclei and their jet-fed lobes. For astrophysical conclusions to be successfully drawn, surveys must be as uniform, unbiased, and well-understood as possible.

A simple attribute of a survey is its flux density calibration accuracy: how closely do the reported flux densities reflect the real flux densities of those sources in our sky? A survey with an overall systematic bias toward higher or lower flux densities will misrepresent the luminosities of distant objects; a survey with position-dependent flux density calibration issues will misreport the flux densities of individual objects; a survey with coherent position-dependent flux density calibration issues can lead to misleading measurements for large groups of sources, or bias results for a complementary small-area survey. Where possible, flux density calibration should be uniform, or at least known.

Another important aspect of a sky survey is its completeness. \cite{1998AJ....115.1693C} explore in detail the trade-offs between survey resolution and sensitivity. Scientific uses of surveys require samples limited by total flux density, not image brightness (``peak flux density"), but it is the latter limit that is determined by the instrument configuration. Only relatively low-resolution surveys can produce the desired uniformly low flux density limit. \cite{1998AJ....115.1693C} point out that follow-up observations can increase the resolution on a detected source, but a missed source is lost forever, so completeness is actually the most fundamental requirement of a survey. Note, however, that with a lower resolution comes a higher confusion limit, which once reached, can only be surpassed by higher-resolution observations.

A large-scale 150\,MHz, 16-MHz bandwidth, continuum survey from the Giant Metrewave Radio Telescope \citep[GMRT;][]{1991CuSc...60...95S} was conducted between April 2010 and March 2012 by the Tata Institute of Fundamental Research (TIFR), and the raw data were made available as the TIFR GMRT Sky Survey (TGSS\footnote{http://tgss.ncra.tifr.res.in/}) via the GMRT archive\footnote{https://naps.ncra.tifr.res.in/goa/mt/search/basicSearch}. This is the highest-resolution low-frequency sky survey yet conducted, but until 2016, only a small number of data products had been published \citep[e.g. ][]{2011ApJ...736L...8B,2012MNRAS.423.1053G,2014A+A...562A.108S,2014MNRAS.443.2824K}. \cite{2017A+A...598A..78I} published an Alternative Data Release of the TGSS (TGSS-ADR1), reducing the data using the \textsc{SPAM} package, a set of AIPS-based data reduction scripts in Python that includes direction-dependent calibration and imaging. This data release covers 36,900 square degrees over $-53^\circ < \mathrm{Declination} < +90^\circ$, with a median RMS noise below 3.5\,mJy\,beam$^{-1}$ and an approximate resolution of $25"\times25"$. Using a detection limit of 7-sigma, the TGSS-ADR1 catalogue comprises 0.62\,Million radio sources with an astrometric accuracy of better than 2" in RA and Dec. The survey has problems with completeness, as it appears to be missing objects which should be detected at high S/N (see~\Sect~\ref{sec:gleaming}).

Another recent low-frequency sky survey was conducted by the Murchison Widefield Array \citep[MWA;][]{2013PASA...30....7T} over the wider band of 72--231\,MHz, from August~2013 to July~2016: the GaLactic and Extragalactic All-sky MWA survey \citep[][]{2015PASA...32...25W}, hereafter referred to as GLEAM. Due to the short ($<3$\,km) baselines of the MWA, the survey resolution is limited to $2'$ at 200\,MHz. Careful source-extraction techniques were used by \cite{2017MNRAS.464.1146H} to produce a catalogue comprising 20~flux density measurements across the full bandwidth for over 0.3\,Million radio sources over $-90^\circ < \mathrm{Declination} < +30^\circ$, excluding regions near the Galactic Plane, the Magellanic Clouds, and Centaurus~A. The median RMS noise of the catalogue is 9\,mJy\,beam$^{-1}$, dominated by sidelobe confusion, and the astrometric uncertainty is about 3" in RA and Dec. With a range of baselines sampling scales from $2'$--15$^\circ$, the completeness of GLEAM is well-understood and measured \citep{2017MNRAS.464.1146H}.

Both surveys estimate their overall flux density calibration accuracy at better than 10\,\%, but this work finds that when their flux density scales are compared, there are position-dependent flux density scale variations of order 15\%, and these are primarily in the TGSS-ADR1. Resolution is, of course, essential, for understanding the morphology of objects and for cross-matching with other wavelengths -- particularly optical and IR, in order to find hosts for radio jets. This work therefore aims to improve the flux density calibration of TGSS-ADR1 by using the more uniformly calibrated GLEAM, for the area $-90^\circ < \mathrm{Declination} < +30^\circ$, with the aim of producing a better-calibrated high-resolution catalogue for future work. In particular, this helps meet the other important goal of surveys, to be complete, since together GLEAM and TGSS offer a powerful combination of completeness and resolution.

\Sect~\ref{sec:gleaming} introduces the catalogues in more detail, and motivates the selection of GLEAM as the baseline catalogue for bootstrapping the flux density calibration. \Sect~\ref{sec:correcting} describes the derivation and application of the flux density scale correction; \Sect~\ref{sec:results} describes the resulting catalogue. \Sect~\ref{sec:conclusions} concludes the work with recommendations for use of the catalogue, and potential applications.

\section{Flux calibration accuracy of GLEAM and TGSS-ADR1}\label{sec:gleaming}

A third catalogue which can be used to semi-independently test the flux density calibration accuracy of GLEAM and TGSS-ADR1 is the Molonglo Reference Catalogue at 408\,MHz \citep[MRC;][]{1981MNRAS.194..693L,1991Obs...111...72L}, which covers $-90^\circ < \mathrm{Declination} < +19^\circ$, containing 12,141 discrete sources with $S>0.7$\,Jy
at a resolution of $2\farcm62$ by $2\farcm86\sec(\mathrm{Dec}+35\fdg5)$. GLEAM is not entirely independent from the MRC, because in conjunction with the VLA Low-frequency Sky Survey Redux at 74\,MHz \citep[VLSSr; ][]{2014MNRAS.440..327L} and the NRAO VLA Sky Survey \citep[NVSS; ][]{1998AJ....115.1693C} at 1.4\,GHz, it was used to determine MWA primary beam corrections during the flux density scaling of GLEAM. However, this flux density calibration was derived by integrating over at least 10~fields-of-view from VLSSr, MRC, and NVSS, due to the large field-of-view of the MWA. This acts to smooth out any potential contribution from flux density scale variations from these surveys. If there are still flux density scale variations on scales $<30^\circ$ in MRC, they will be visible in the analysis presented in this paper. We also note that there are no other low-frequency wide-area surveys covering the equatorial region.

\begin{figure*}
    \centering
    \includegraphics[width=\textwidth]{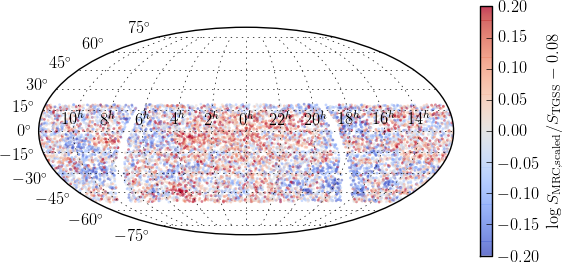} 
    \includegraphics[width=\textwidth]{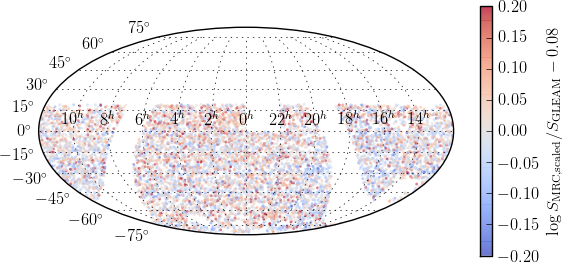} 
    \caption{The log flux density ratios of TGSS-ADR1 (top) and GLEAM (bottom) with respect to MRC, the source flux densities of which were scaled by power-law SEDs with $\alpha=-0.82$ to match the central frequencies of TGSS-ADR1 and GLEAM, 150 and 200\,MHz, respectively.
The ratios have been corrected by the median offset, $0.08$, for clarity.
The RMS of the log flux density ratios is 17\% and 9\% for TGSS-ADR1 and GLEAM, respectively.}
    \label{fig:mrcratio}
\end{figure*}

As the results of this comparison will not be used to create perform any corrections, merely to assess the magnitude of the direction-dependent flux density scale errors in each survey, we perform simple distance-based cross-matching within a radius of 25" between: GLEAM and MRC, and TGSS-ADR1 and MRC. To scale the flux densities of MRC to the GLEAM and TGSS frequencies of 200 and 150\,MHz, we apply a simple power-law SED of $S\propto\nu^\alpha$, setting $\alpha$ to $-0.82$, the median spectral index of reasonably bright radio sources between 74 and 1400~MHz \citep{2014MNRAS.440..327L}. The resulting log \footnote{This, and all further logarithms, are in base 10.} flux density ratios are plotted in \Fig~\ref{fig:mrcratio}.

The median flux density ratios of MRC to both GLEAM and TGSS-ADR1 are about 1.2, i.e. a slightly flatter spectral index of $\alpha\approx-0.6$ is favoured over the canonical $-0.82$. This is due to the inclusion of fainter sources than those used by \cite{2014MNRAS.440..327L}; these faint sources tend to have flatter spectral indices. There are potentially also more sources with low-frequency turnovers, which make up a few per~cent of the low-frequency population at these flux densities \citep{2017ApJ...836..174C}. The effect reduces to within the flux density scale errors of the two surveys at higher flux density cut-offs of several Jy, but such a limit renders position-dependent effects much less visible.

More importantly, however, the log flux density ratio of MRC to GLEAM is visibly flat with respect to sky position, with an RMS of 9\,\%, while that of MRC to TGSS-ADR1 varies in a coherent position-dependent way, with an RMS of 17\,\%. This indicates that the flux density scale of TGSS-ADR1 is less uniform than that of GLEAM, and correcting it to the GLEAM flux density scale would make it more uniform.

Note also that both TGSS-ADR1 and GLEAM are slightly brighter near the Galactic Centre (RA$\approx17.5^\mathrm{h}$, Dec$\approx-27^\circ$) than MRC, manifesting as a slightly more negative log flux density ratio. This area is difficult to image with all three instruments, due to the presence of bright large-scale Galactic synchrotron emission, so it is unclear whether this is a problem with MRC, or both TGSS-ADR1 and GLEAM. In either case, since the feature is visible in both TGSS-ADR1 and GLEAM, any correction will only minimally affect it.

\begin{figure*}
    \centering
    \includegraphics[width=\textwidth]{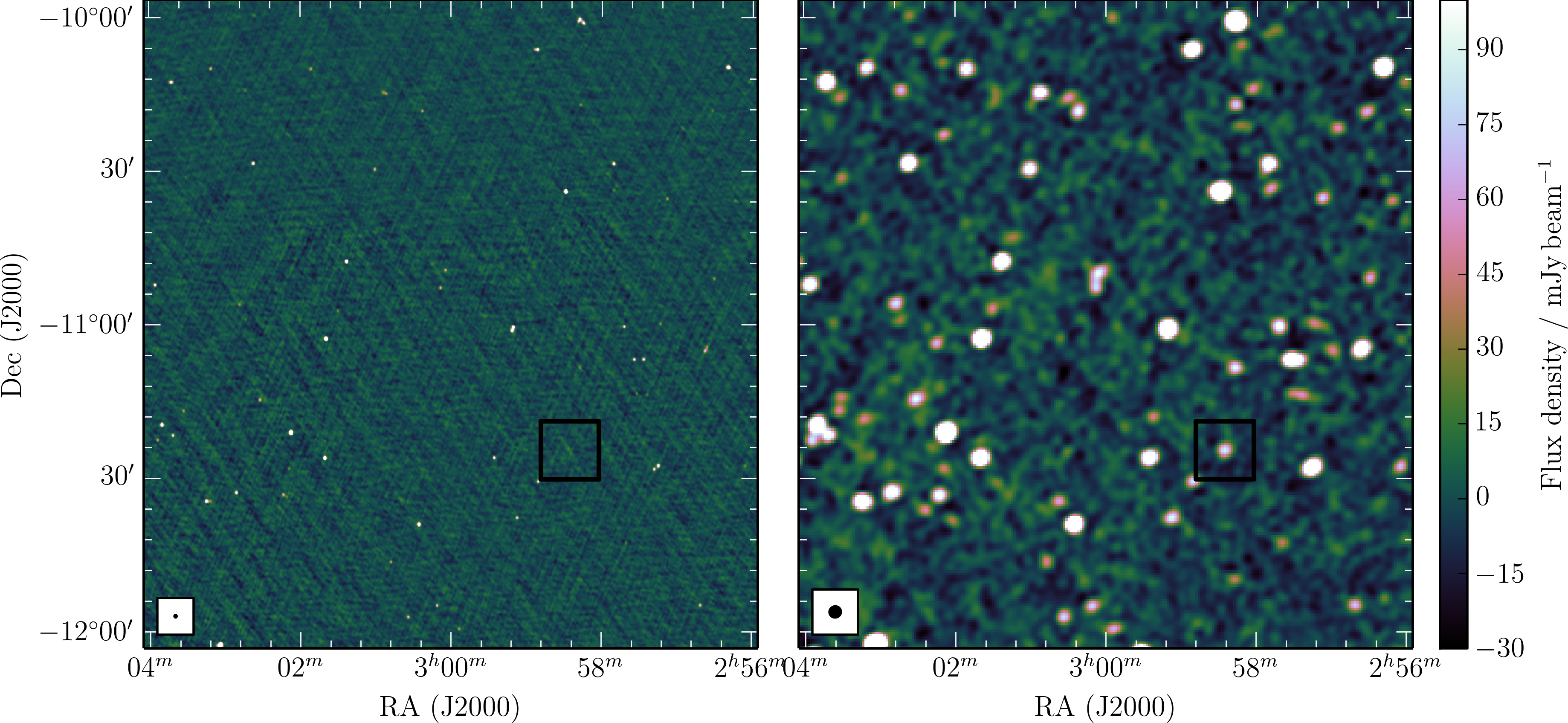} 
    \includegraphics[width=\textwidth]{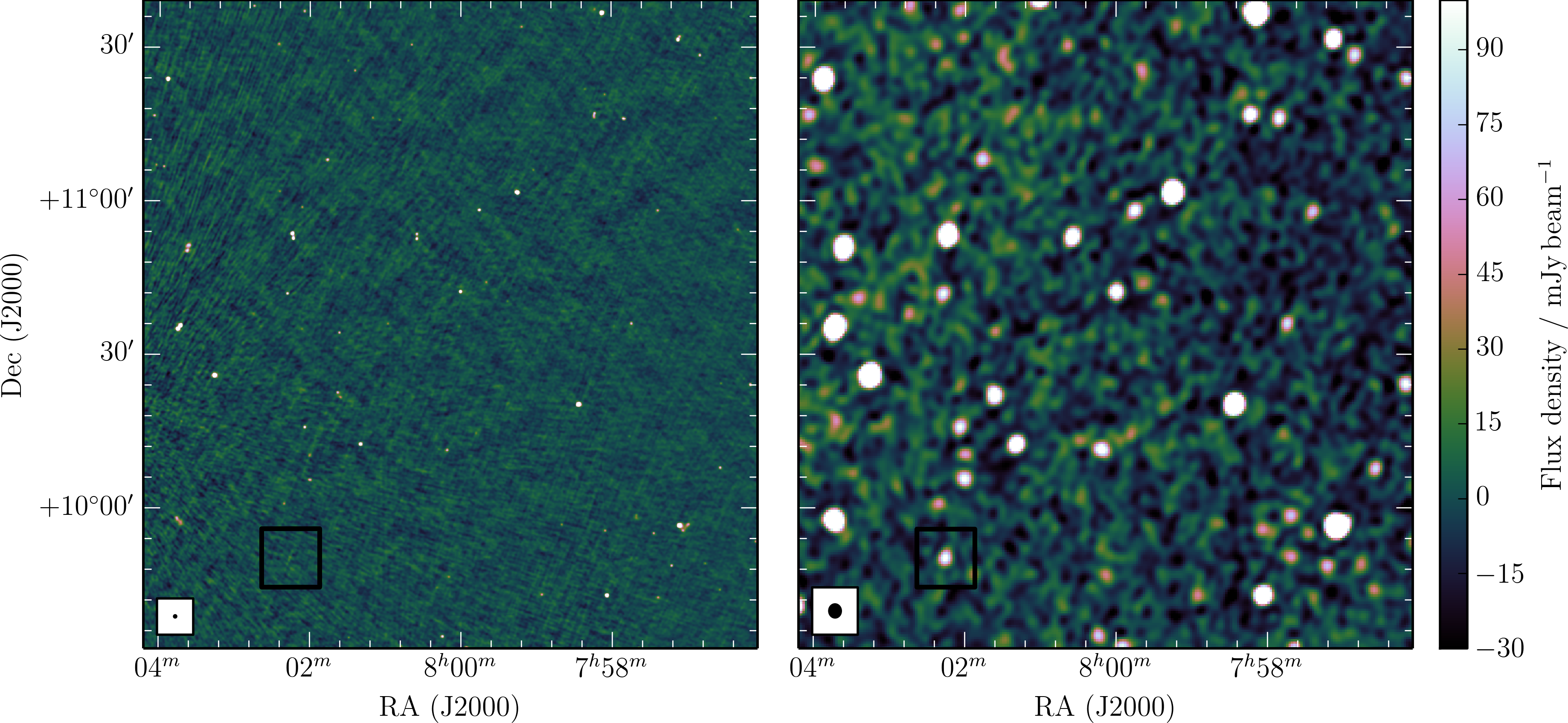} 
    \caption{Two regions selected from both TGSS-ADR1 (left; $\nu=150$\,MHz, $\Delta\nu=16$\,MHz; $\mathrm{RMS}\approx4$\,mJy\,beam$^{-1}$) and GLEAM (right; $\nu=200$\,MHz, $\Delta\nu=60$\,MHz; $\mathrm{RMS}\approx9$\,mJy\,beam$^{-1}$) to illustrate image quality, resolution, and surface brightness sensitivity in regions of differing TGSS-ADR1 flux density scales. In each figure, a source present in GLEAM (at $>10\sigma$) but absent in TGSS-ADR1 (at $>20\sigma$) has been highlighted with a black square in each panel, to indicate the completeness issues in TGSS-ADR1. The point spread function for each image is shown as a filled ellipse in the bottom left of each panel. Top: RA~03$^\mathrm{h}$, Dec~$-11^\circ$: TGSS-ADR1 is $\approx20$\% low relative to MRC; highlighted source is GLEAM~J025825-112435 ($S_\mathrm{200MHz}=120$\,mJy); bottom: RA~$08^\mathrm{h}$, Dec~$+10\fdg5$: TGSS-ADR1 is as consistent with MRC as GLEAM; highlighted source is GLEAM~J080215+095017 ($S_\mathrm{200MHz}=117$\,mJy) .}
    \label{fig:images}
\end{figure*}

We briefly consider the image quality of TGSS-ADR1 and GLEAM in regions of different flux density scales, in order to check that a simple flux density scaling is the only correction necessary to reconcile the two catalogues. We select $2\times2^\circ$ areas around RA~03$^\mathrm{h}$, Dec~$-11^\circ$ (TGSS-ADR1 is ~20\% low relative to MRC), and RA~$08^\mathrm{h}$, Dec~$+10\fdg5$ (TGSS-ADR1 is as consistent with MRC as GLEAM). The images are shown in \Fig~\ref{fig:images}: there is no noticeable change in image quality between the regions for either survey. We conclude that a simple flux density correction is adequate for matching the two surveys for typical areas of sky.

Some high-significance sources are only seen in GLEAM: two examples are highlighted in \Fig~\ref{fig:images}. We measure a spatial density of about 0.15~missing sources per square degree. This was determined by selecting a region of high completeness in GLEAM (RA $0^\mathrm{h}$--$5^\mathrm{h}$, Dec $-40^\circ$--$0^\circ$) and selecting only sources that: are isolated (no other GLEAM source within 300''); are above a flux density at which the catalogue shows 90\% completeness ($S>70$\,mJy); and are unresolved ($S_\mathrm{int}/S_\mathrm{peak} <= 1.2$). These sources were then cross-matched against TGSS-ADR1, selecting only sources which appear in GLEAM but do \textbf{not} have a match in TGSS-ADR1 within five arcmin. Finally, we excluded any sources in regions with a TGSS-ADR1 RMS noise of $>5$\,mJy. This leaves 452~sources in this area, giving the aforementioned spatial density. These sources are not present in TGSS-ADR1 with a significance of $S_\mathrm{TGSS}/\mathrm{RMS}_\mathrm{TGSS}>12$.

These are extremely unlikely to be transient objects, as this would be inconsistent with long monitoring campaigns from other low-frequency instruments \citep[see e.g.][]{2016MNRAS.456.2321S}.
In some areas, the noise in the TGSS-ADR1 images appears less Gaussian, perhaps indicating calibration artefacts. However, we attempt to exclude these regions with the final TGSS RMS cut. A potential cause is a bias in TGSS-ADR1 against low surface brightness objects, due to its higher resolution and its filtering of short ($<200\lambda$) baselines. The ultimate cause will likely be addressed in future TGSS data products; in the meantime, a combination of TGSS-ADR1 and GLEAM provides a more complete picture of the southern sky at these frequencies, but reconciling the flux density scales is necessary for most uses.

\section{Creating a corrected catalogue}\label{sec:correcting}

\subsection{Flux scale sample}\label{ssec:sample}

In order to perform a robust modification of the flux density scale of TGSS-ADR1, a representative sample of radio galaxies avoiding systematic biases must be chosen. Sources must match the following criteria:
\begin{itemize}
\item{Isolated: To avoid cross-matching single components of doubles resolved in TGSS-ADR1, but unresolved in GLEAM, TGSS-ADR1 sources cannot lie within a $1'$ radius of another TGSS-ADR1 source;}
\item{Bright: Both GLEAM and TGSS-ADR1 integrated (total) flux density $S_{150\mathrm{MHz}}>400$\,mJy;}
\item{Spectrally simple: GLEAM reports a non-null $\alpha$, indicating that a power-law spectrum is a good fit (reduced $\chi^2<1.93$);}
\item{Unresolved: Both the GLEAM and TGSS-ADR1 ratio of integrated to peak flux density is $<1.2$.}
\end{itemize}

This selects 47,984~sources from TGSS-ADR1 and 51,993~sources from GLEAM. Crossmatching using a radius of $25"$ forms a catalogue of 23,796~sources. Varying the crossmatching radius between 10 and $120"$ yields only $\approx5$\% differences in the number of sources cross-matched, due to the downselection criteria employed; $25"$ is a compromise between matching the most sources and avoiding false positives.

To bring the two catalogues to the same frequency (150\,MHz), we take advantage of the wide bandwidth of GLEAM, which was used by \citet{2017MNRAS.464.1146H} to fit power-law spectra to all detected sources. For those sources with a reasonable fit (a reduced $\chi^2<1.93$, i.e. a 99\% probability of the SED being a power-law), the GLEAM catalogue reports $S_\mathrm{200\,MHz,fitted}$ and $\alpha$: the 200\,MHz GLEAM flux density and spectral index calculated by fitting over all 20~7.68-MHz sub-bands of GLEAM. Any sources too faint or not statistically well-described by a power law do not have these values, and are not used in this analysis.

The choice of 200\,MHz as the reported frequency for this fitted flux density is fairly arbitrary, and it is trivial to rescale to 150\,MHz using the accurate low-frequency $\alpha$ via $S_\mathrm{150\,MHz,GLEAM} = S_\mathrm{200\,MHz,fitted}\times(\frac{150\,\mathrm{MHz}}{200\,\mathrm{MHz}})^\alpha$. This is more accurate than using the value at the closest-matching sub-band centred on 151\,MHz, which has higher noise and is only a single measurement, instead of a fit to twenty. From the crossmatch of the two catalogues, we can then calculate the spatial distribution of the log flux density ratio, $\log{\frac{S_\mathrm{150\,MHz,GLEAM}}{S_\mathrm{150\,MHz,TGSS}}}$.

The distribution of the log flux density ratio is shown in the top panel of \Fig~\ref{fig:ratio} (other panels are described in \Sect~\ref{ssec:rbf}). The coherent patches of similar flux density scales correspond to different TGSS observing nights with the GMRT. The observing strategy was to perform 3-minute observations of target fields laid out in a hexagonal configuration on the sky. Fields were arranged in $\approx5^\circ\times5^\circ$~patches, and each patch was observed for 2--3 hours, encompassing 60--70 target fields. A different patch would then be selected, and the same process repeated. Phase calibration scans occurred once per hour, and flux calibration scans were performed at the beginning and end of each night, although typically only one would be used in the final flux density calibration process.

As an example, two black boxes highlight observations taken on 2010-12-14 (project code 19\_043, night 5142). To obtain these regions, pointing centres of the observations were downloaded from the GMRT online archive\footnote{https://naps.ncra.tifr.res.in/goa/mt/search/basicSearch}, and the enclosing sky area measured. This was then expanded by a border of 1\fdg5, corresponding to the GMRT 150\,MHz primary beam full-width-half-maximum. Within both patches on the same night, the flux density scale is close to uniform, due to the use of the same flux/bandpass calibrator.
Since no secondary calibrators are used, if the initial flux calibration is incorrect, it will be applied consistently throughout the night. The calibration scan may not be correct due to delay jumps in the antenna gains which will be fixed in a future TGSS data release \citep{2017A+A...598A..78I}.

The increased signal-to-noise of these patches compared to the MRC cross-match shown in \Fig~\ref{fig:mrcratio} is due to the use of the much more accurate GLEAM spectral index $\alpha$ measurements than a canonical $-0.82$. There is no correspondence with the observing parameters of GLEAM \citep{2015PASA...32...25W} or MRC \citep{1981MNRAS.194..693L}, both of which were performed as long (8--12-hour)~drift scans.

\subsection{Radial basis function fitting}\label{ssec:rbf}

The value of a radial basis function is dependent only on the distance from the origin, and a sum of multiple radial basis functions with different centres can be used as an approximation to a varying surface. The \textsc{python} package \textsc{scipy} \citep{scipy} includes a set of interpolation functions, including \texttt{interpolate.Rbf}, which we use here to fit the spatial distribution of log flux density ratios. Using the simplest version of the function, we specifiy a linear radial basis function, i.e. $\phi(r) = r$, with a smoothness of $10^\circ$, the smallest believable spatial scale of the variations. (The data can be arbitrarily overfit with very low smoothness values, but this will mean a poorer extrapolation to other sources in the field.)

\begin{figure*}
    \centering
    \includegraphics[width=\textwidth]{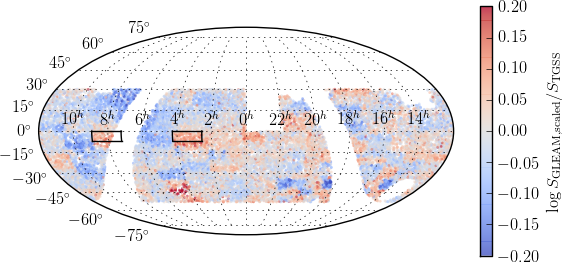} 
    \includegraphics[width=\textwidth]{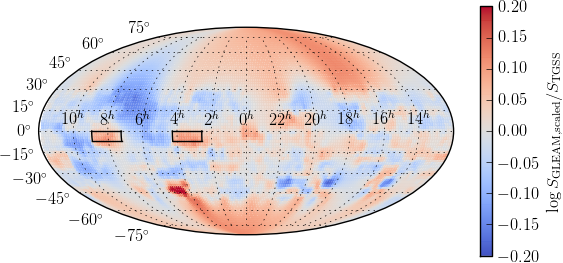} 
    \includegraphics[width=\textwidth]{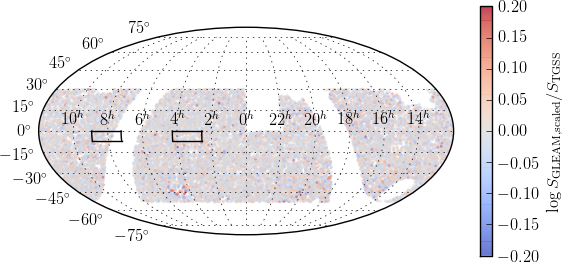} 
    \caption{The log flux density ratio of GLEAM to TGSS-ADR1 before correction (top), and after (bottom). The middle panel shows the model, at a $1\arcdeg$ grid resolution. The two highlighted areas show the observing that took place on 2010-12-14, as an example of typical GMRT observing strategies used during TGSS. Within the observations taken on the same night, the flux density scale is close to uniform, and this is accounted for in the model fit.}
    \label{fig:ratio}
\end{figure*}

The resulting fit is shown in the middle panel of \Fig~\ref{fig:ratio}, while the difference between the fit and the data is shown in the lowest panel. The remaining residuals are noise-like and no longer correlate with the observing parameters of the TGSS.

\section{Results}\label{sec:results}

\begin{figure*}
    \centering
    \includegraphics[width=\textwidth]{TGSS_MRC_fluxratio.png} 
    \includegraphics[width=\textwidth]{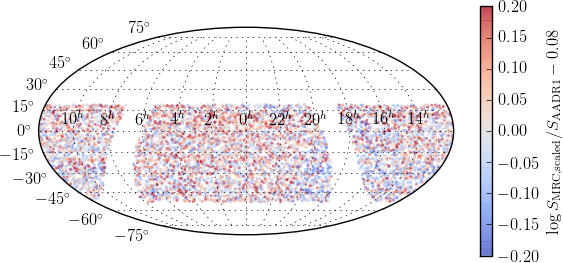} 
    \caption{The log flux density ratios of TGSS-ADR1 and TGSS-RSADR1 with respect to MRC, the flux densities of which were scaled by power-law SEDs with $\alpha=-0.82$ to 150\,MHz.
The ratios have been corrected by the median offset, $0.08$, for clarity.
The RMS of the log flux density ratios is 17\% and 12\% for TGSS-ADR1 and TGSS-RSADR1, respectively.}
    \label{fig:mrcaadr1}
\end{figure*}

Applying the fit to the TGSS-ADR1 catalogue results in a modified version: for each source, the values in the catalogue of
Total flux [density], Error on total flux [density], Peak flux [density], Error on peak flux [density], and RMS noise are multiplied by the flux scale ratio calculated at the position of the source.
We can check the impact of the position-dependent flux density scale modification by repeating the MRC comparison detailed in \Sect~\ref{sec:gleaming}. The resulting log flux density ratio is plotted in \Fig~\ref{fig:mrcaadr1}; the median is 1.2, as expected, and the RMS has decreased from 17\% to 12\%, not quite as uniform as GLEAM, but closer. Most importantly, the position-dependent flux density scale patchiness is mostly erased; the variations are now randomly distributed. The slight over-brightness is still visible near the Galactic Centre, which is unsurprising, as GLEAM shows the same feature (see \Sect~\ref{sec:gleaming}).
The small regions not included in the GLEAM extragalactic catalogue due to adverse ionospheric conditions and the presence of Centaurus~A (see Table~3 of \citealt{2017MNRAS.464.1146H}) are uniform in the corrected figure, so we assume the interpolation has been reasonably successful over these areas (i.e. the flux density scale variations were not large or discontinuous in the first place). The Galactic plane is a larger area missing from GLEAM, and therefore any interpolated flux density scale ratios therein are less accurate, and any corrections North of Declination~$30^\circ$ are purely extrapolations.

We therefore apply spatial cut-offs to the catalogue of $\mathrm{Declination}\leq+30^\circ$, $|b|\geq10^\circ$, reducing the sky area covered to 18,800\,deg$^2$. After applying these cuts, 383,589~TGSS-ADR1 sources remain, which we release as TGSS-RSADR1, a Rescaled Subset of Alternative Data Release~1. The modified catalogue contains the original TGSS-ADR1 column headings, plus an additional flux density scaling ratio to indicate the correction value that was applied. An example few lines from the catalogue are shown in \Tab~\ref{tab:catalogue}. We note that TGSS-RSADR1 is on the flux density scale of \citet{Baars1977}, like GLEAM.

The catalogue is available from the author on request.

\begin{sidewaystable}
\tiny \centering
\caption{Random sample of 12~entries from the TGSS-RSADR1 source catalog. Columns 1--5, 8--10, 12, and 13, are identical to those published by \cite{2017A+A...598A..78I}. Columns 6, 7, and 11 are those given in \cite{2017A+A...598A..78I} but multiplied by the value given in column~14, the flux density scale ratio for that source as determined by the fitting process described in \Sect~\ref{ssec:rbf}. The full catalogue of 383,589~sources is available via CDS.\label{tab:catalogue}}
\begin{tabular}{cccccccccccccc}
\hline
\hline
  \multicolumn{1}{c}{Source\_name} &
  \multicolumn{1}{c}{RA} &
  \multicolumn{1}{c}{$\sigma_\mathrm{RA}$} &
  \multicolumn{1}{c}{Dec} &
  \multicolumn{1}{c}{$\sigma_\mathrm{Dec}$} &
  \multicolumn{1}{c}{$S_\mathrm{total}$} &
  \multicolumn{1}{c}{$S_\mathrm{peak}$} &
  \multicolumn{1}{c}{Maj} &
  \multicolumn{1}{c}{Min} &
  \multicolumn{1}{c}{PA} &
  \multicolumn{1}{c}{RMS noise} &
  \multicolumn{1}{c}{Code} &
  \multicolumn{1}{c}{Mosaic} &
  \multicolumn{1}{c}{$S$ scale ratio} \\
  (IAU) & [$^\circ$] & [$"$] & [$^\circ$] & [$"$] & [mJy] & [mJy\,beam$^{-1}$] & [$"$] & [$"$] & [$^\circ$] & [mJy\,beam$^{-1}$] & [S/M/C] & & \\
  (1) & (2) & (3) & (4) & (5) & (6) & (7) & (8) & (9) & (10) & (11) & (12) & (13) & (14) \\
\hline
  TGSSRSADR J193757.7-082457 & 294.4908 & 2.3 & $-8.4161$ & 2.7 & $75.2\pm9.2$ & $38.9\pm5.4$ & $41.7\pm4.3$ & $32.8\pm2.9$ & $-6.7\pm18.6$ & 3.4 & S & R60D27 & 1.026\\
  TGSSRSADR J225418.8-514856 & 343.5785 & 2.3 & $-51.81583$ & 5.4 & $59.0\pm9.0$ & $39.5\pm5.7$ & $80.0\pm11.9$ & $34.4\pm2.6$ & $-0.0\pm6.6$ & 4.4 & C & R69D02 & 1.073\\
  TGSSRSADR J065146.3+242334 & 102.94319 & 2.0 & $24.39278$ & 2.0 & $286.5\pm30.1$ & $208.4\pm21.3$ & $31.3\pm0.8$ & $24.3\pm0.5$ & $82.8\pm4.1$ & 4.3 & M & R21D50 & 0.697\\
  TGSSRSADR J121142.5+003749 & 182.92741 & 2.1 & $0.63049$ & 2.1 & $43.2\pm6.0$ & $40.8\pm4.8$ & $27.7\pm1.8$ & $25.1\pm1.5$ & $51.9\pm26.3$ & 2.5 & S & R37D34 & 1.008\\
  TGSSRSADR J120443.5-475111 & 181.1816 & 2.1 & $-47.85309$ & 2.0 & $426.2\pm43.3$ & $286.8\pm28.9$ & $76.0\pm1.3$ & $29.3\pm0.2$ & $-6.6\pm0.7$ & 3.3 & M & R37D03 & 0.957\\
  TGSSRSADR J045939.8-320308 & 74.91595 & 2.0 & $-32.05223$ & 2.2 & $160.4\pm18.9$ & $142.6\pm15.4$ & $44.0\pm2.3$ & $26.4\pm0.8$ & $1.2\pm4.0$ & 5.9 & S & R16D11 & 0.812\\
  TGSSRSADR J012510.4+124151 & 21.29345 & 2.0 & $12.69752$ & 2.0 & $375.3\pm38.2$ & $204.9\pm20.7$ & $36.9\pm0.7$ & $25.8\pm0.4$ & $-72.1\pm2.3$ & 3.0 & M & R05D42 & 0.976\\
  TGSSRSADR J024056.1+251148 & 40.23411 & 2.0 & $25.19667$ & 2.0 & $272.9\pm27.7$ & $262.8\pm26.4$ & $25.8\pm0.3$ & $25.2\pm0.3$ & $-7.9\pm20.2$ & 2.9 & S & R09D49 & 0.872\\
  TGSSRSADR J083856.2-111019 & 129.73448 & 2.4 & $-11.17205$ & 2.7 & $28.7\pm7.0$ & $27.3\pm4.6$ & $29.9\pm4.4$ & $25.6\pm3.2$ & $0.5\pm37.9$ & 3.7 & S & R27D25 & 1.0624\\
  TGSSRSADR J122110.0-195854 & 185.29182 & 2.0 & $-19.98185$ & 2.0 & $191.5\pm19.9$ & $158.0\pm16.1$ & $39.5\pm1.0$ & $24.9\pm0.4$ & $-1.3\pm2.2$ & 3.3 & S & R38D19 & 1.167\\
  TGSSRSADR J114822.8-462756 & 177.09515 & 2.1 & $-46.46563$ & 3.1 & $69.6\pm9.0$ & $56.6\pm6.5$ & $66.3\pm5.6$ & $28.4\pm1.1$ & $6.2\pm4.2$ & 3.5 & S & R36D04 & 0.991\\
  TGSSRSADR J012738.7-032702 & 21.91146 & 2.8 & $-3.4508$ & 2.7 & $35.5\pm5.6$ & $22.1\pm3.6$ & $36.4\pm5.1$ & $30.1\pm3.6$ & $54.9\pm31.0$ & 2.7 & S & R05D30 & 1.059\\
\hline\end{tabular}
\end{sidewaystable}

\section{CONCLUSION}\label{sec:conclusions}

We have produced a modified version of the TGSS-ADR1 source catalogue that is consistent in flux density scale with the first-year GLEAM extragalactic radio source catalogue \citep{2017MNRAS.464.1146H}, thereby reducing position-dependent flux density scale variations.

The improved flux density scale alignment with GLEAM makes TGSS-RSADR1 useful for several applications:
\begin{itemize}
\item{Combining GLEAM and TGSS-RSADR1 to form a more complete sample of sources across the overlapping region, allowing better ensemble statistics of radio galaxies;}
\item{Utilising the morphological resolving power of TGSS with the spectral coverage of GLEAM to better-characterise many sources;}
\item{Looking for extended sources with higher GLEAM flux densities than TGSS-RSADR1, to find low surface brightness emission, a signature of radio relics and haloes, as well as extended radio galaxies;}
\item{Creating better sky models for calibrating LOFAR, the extended MWA, SKA1\_LOW, HERA, and other upcoming low-frequency instruments;}
\item{Checking the flux density calibration of a second Alternative Data Release, which will improve on the phase calibration and consistency of ADR1 by using further observations and modified data reduction (Intema et al. in prep).}
\end{itemize}

We request that any users of this catalogue cite both this paper and \cite{2017A+A...598A..78I}, the TGSS-ADR1 release on which this work is significantly based.

\begin{acknowledgements}
The author would like to thank Huib Intema for producing the TGSS-ADR1 catalogue and being amenable to the publication of this modified version. His useful comments, as well as those of Melanie Johnston-Hollitt, Lister Staveley-Smith, and the anonymous referees, improved the paper.
\end{acknowledgements}

\begin{appendix}

\end{appendix}

\bibliographystyle{apj}
\bibliography{refs}

\end{document}